\documentclass[12pt,preprint]{aastex}

\shorttitle{Non-equilibrium Ionization Plasma During Limb Flare}
\shortauthors{Imada}

\begin{document}

\title{NON-EQUILIBRIUM IONIZATION PLASMA DURING LARGE SOLAR LIMB FLARE OBSERVED BY HINODE/EIS}

\author{
S. \textsc{Imada},\altaffilmark{1} 
}
  
\altaffiltext{1}{ Institute for Space-Earth Environmental Research (ISEE), Nagoya University, Furo-cho, Chikusa-ku, Nagoya 464-8601, Japan}

\begin{abstract}
This study on plasma heating considers the time-dependent ionization process during a large solar flare on September 10, 2017, observed by Hinode/EIS. The observed \ion{Fe}{24} / \ion{Fe}{23} ratios increase downstream of the reconnection outflow, and they are consistent with the time-dependent ionization effect at a constant electron temperature Te = 25 MK. Moreover, this study also shows that the non-thermal velocity, which can be related to the turbulent velocity, reduces significantly along the downstream of the reconnection outflow, even when considering the time-dependent ionization process.
\end{abstract}

\keywords{Sun: photosphere---Sun: flow---Sun: solar cycle}

\section{INTRODUCTION}
Magnetic reconnection has been recognized as one of the key mechanisms for heating and bulk acceleration of space plasmas. This energy conversion mechanism is not limited to the solar atmosphere \citep[e.g.,][]{pne}, but was also observed in the Earth's magnetosphere \citep[e.g.,][]{hon,oie,ima2005,ima2007,ima2011} and in laboratory \citep[e.g.,][]{ono,yam,ji}, as well as in other space plasmas. The solar atmosphere is an excellent space laboratory for magnetic reconnection study, as it allows large-scale observation of magnetic reconnection. 

Over the past several decades, many studies have been conducted to explain the conversion of magnetic energy into plasma energy during solar flares, and various models have been proposed. It is now widely believed that magnetic reconnection (as per the so-called CSHKP model) is the fundamental energy conversion mechanism of flares \citep[][]{car,stu,hir,kop}. Modern telescope observations confirm many of the features predicted by the reconnection model (the cusp-like structure: \cite{tsu}; high-energy electron acceleration: \cite{mas}; chromospheric evaporation: \cite{ter}; \cite{ima2008,ima2015}; reconnection inflow: \cite{yok}; outflows: \cite{har,liu,ima2013}; plasmoid ejection: \cite{ohy}; CMEs: \cite{sve}; \cite{ima2007b,ima2011b}). 
As predicted by the CSHKP model, magnetic reconnection occurs above the flare arcade. To date, many observations have been made on the solar limb to confirm the presence of high-temperature and high-speed plasma flows produced by magnetic reconnection above flare arcades. Limb observation has the advantage that the height information can be clearly determined. For example, Super Arcade Downflows (SADs) with reconnection outflow predicted from the flare model are observed by corona imagers. During the gradual phase of flares, dark voids and sometimes bright features of X-rays move from the high corona toward the solar surface (downward) at an apparent velocity of a few 100 km s$^{-1}$, \cite[e.g.,][]{mck,mck2,sav}. On the other hand, equivalent spectroscopic observations are extremely rare, and only a few spectroscopic limb observations have reported the presence of hot fast flows above the flare arcade \cite[e.g.,][]{inn,wan,ima2013}. Most of these studies were based on a single slit position above the flare arcades, so spatial information on hot fast flows was not available. Recently, \cite{war} reported on the structure and evolution of a current sheet formed above the flare arcade. They found that plasma heating seems to occur downstream of the reconnection outflow and reaches temperatures of approximately 25 MK with the ionization equilibrium assumption. 

In the solar corona, plasma is believed to be in thermal equilibrium because of the presence of a weak Coulomb collision. Many studies have discussed the plasma dynamics in a solar corona based on the thermal equilibrium assumption. Most phenomena observed in the solar corona can be explained by the assumption that the temporal resolution is not sufficient to resolve the non-equilibrium conditions. \cite{ima2011c} pointed out that ionization cannot reach equilibrium in the magnetic reconnection region because of its fast flow and rapid heating. In fact, the timescale for ionization ($\sim$100 s) is comparable to the Alfv\'en timescale ($\sim$100 s) in the magnetic reconnection region. Therefore, it is important to consider time-dependent ionization processes when interpreting observations of the magnetic reconnection region. In this letter, I will discuss plasma heating considering the time-dependent ionization process during the X8.3 flare on September 10, 2017, which was reported by \cite{war}.
  
\section{DATA}
On September 10, 2017, a large solar flare (GOES X8.3, peak time 16:06) occurred in the northwest solar limb (30$^\circ$ N, 90$^\circ$ W). 
Several studies have successfully captured the characteristics of flare-associated phenomena, such as temperature \citep[][]{war}, flows \citep[][]{che,lon}, non-thermal velocity  \citep[][]{li,pol}, and structures in the hot plasma sheet above the post-flare loop.
Elemental abundances in coronal post-flare loops have also been studied by \cite{dos}. Many new findings, such as the presence of microstructures and turbulence inside the plasma sheet, have been discussed with this flare event  \citep[e.g.,][]{cai,fre}. 
The EUV Imaging Spectrometer (EIS) aboard Hinode is a high-spectral/spatial resolution spectrometer aimed at studying dynamic phenomena in the corona \citep{cul}. Hinode EIS observed a flare with a slit scanning mode over a field of 240''$\times$304'' with a 2 arcsec wide slit and 3 arcsec steps between exposures. The exposure time at each position was fixed at 5 s, and the total time for each raster was 535 s. The observation period was from 05:44 to 16:53 UT on September 10, 2017. To process the EIS data, we used the software provided by the EIS team (eis\_prep), which corrects for the flat field, dark current, cosmic rays, and hot pixels. Since in this study I am interested in temperature diagnostics in flare heated plasma, I used only \ion{Fe}{24} 255.10 \AA~and \ion{Fe}{23} 263.76  \AA~data. 

\section{TIME-DEPENDENT IONIZATION}
In order to study plasma heating considering the time-dependent ionization process during a flare, the time evolution of the charge state of iron in the downstream of reconnection outflow was calculated in the same manner as in \cite{ima2011c}. The continuity equations for iron are expressed as follows:
\begin{equation}
\frac{\partial n^{Fe}_i}{\partial t}+\nabla \cdot n^{Fe}_i {\bf v}  = 
n_e\left[n^{Fe}_{i+1} \alpha^{Fe}_{i+1}+ n^{Fe}_{i-1} S^{Fe}_{i-1}-n_i^{Fe}\left(\alpha^{Fe}_{i}+S^{Fe}_{i}\right)\right],
\end{equation}
where $n_i^{Fe}$ is the number density of the {\it i}th charge state of the iron, $\alpha^{Fe}$ represents the collisional and dielectronic recombination coefficients, $S^{Fe}$ represents the collisional ionization coefficients, ${\bf v}$ represents flow velocity, and $n_e$ represents the electron density. The calculation code used in this study is the same as that used by \cite{ima2011c, ima2015}. The ionization and recombination rates were calculated using the work of  \cite{arn1}, \cite{arn2}, and \cite{maz}. Here, it is assumed that all ions and electrons have the same flow speed and temperature in the same location. 

Assuming ionization equilibrium, \cite{war} found that the temperature of the current sheet formed above the flare arcade heated to approximately 20 MK. The highest temperatures were observed in the cusp of the flare arcade, and the temperature in the current sheet decreased with height. This can be interpreted as the plasma being heated along the reconnection outflow. Because the temperature estimated by the ionization equilibrium assumption can be lower than the actual temperature, I calculated the time evolution of the charged state of iron assuming an electron temperature (Te = 20, 25, 30 MK) higher than the temperature estimated in a previous study.
Only \ion{Fe}{23} and \ion{Fe}{24} are shown in the case of Te$=$25 and 30 MK. 
The dotted and dashed lines in Figure 1 show the peak times of \ion{Fe}{23}/\ion{Fe}{24}, respectively.
The electron density with the isothermal assumption was also discussed by \cite{war}. 
They concluded that the density is 10$^{10}$ /cc, which is much larger than the density considered in many previous studies. 
\cite{lon} also discussed the density  of hot flare plasma in the current sheet and concluded that the density is $\sim 10^{10}$ /cc or less.
Time-dependent ionization is typically highly dependent on density.
If the density becomes 10 times higher, then the equilibrium time scale becomes 10 times shorter, and vice versa
Therefore, to avoid overestimating the time-dependent ionization effect, the electron density was assumed to be 10$^{10}$ /cc. 
Figure 1 shows the temporal variation in the charge state of iron. 
The horizontal axis represents the time on a logarithmic scale, and the vertical axis shows the ionization fraction of iron on a logarithmic scale.
The initial temperature is assumed to be 1 MK. 
Ionization reaches equilibrium in approximately 10 s in all cases.
The timescale of the ionization equilibrium is temperature-independent.
The rate of decrease in \ion{Fe}{23} immediately before reaching equilibrium is clearly higher as the temperature rises.
However, the \ion{Fe}{24} does not change significantly after its peak time in all cases.
Note that the time-dependent ionization is not sensitive to the initial temperature of the ionization equilibrium because the ionization process is fast at the beginning.

\section{NON-EQUILIBRIUM IONIZATION PLASMA OBSERVED BY HINODE/EIS}

The time evolution of the charge state of iron downstream of the reconnection outflow discussed in the previous section was applied to the Hinode/EIS observation of the large flare on September 10, 2017. The EIS \ion{Fe}{23} (263.76 \AA) image is shown in Figure 2a. A sheet structure formed above the cusp-shaped structure is observed. The intensity ratio between \ion{Fe}{24} (255.10 \AA) and \ion{Fe}{23} (263.76 \AA) is shown in Figure 2b. The diamonds in Figure 2b represent the height dependence (white dashed line in Figure 2a) of the intensity ratio between \ion{Fe}{24} (255.10 \AA) and \ion{Fe}{23} (263.76 \AA) observed by Hinode/EIS. This result is almost the same as that reported by \cite{war}. The red solid line shows the intensity ratio between Fe XXIII and Fe XXIV considering the time-dependent ionization effect with a constant electron temperature Te = 25 MK. It is assumed that the reconnection point is located at X = 1100 arcsec, and the plasma is assumed to move downstream (toward X = 1000 arcsec) at a rate of 1000 km/s. Therefore, the ionization proceeds as X becomes smaller, and the \ion{Fe}{24} / \ion{Fe}{23} ratio becomes larger. The intensity ratios considering the time-dependent ionization effect with constant electron temperatures Te = 20 and 30 MK are also shown with dashed lines in Figure 2b. The observed ratios (diamonds) seem to be consistent with the results obtained considering the time-dependent ionization effect at a constant electron temperature Te = 25 MK.

To discuss the non-thermal velocity, the line width of the emission line of \ion{Fe}{24} (192.04 \AA) was used. The full width at half-maximum (FWHM), $W_{obs}$, of the observed lines in velocity units, is described by 
\begin{equation}
W_{obs}=\sqrt{W_I^2+4\log{2}\left(\frac{2k_BT_{ion}}{M}+\xi^2\right)},
\end{equation}
where $W_I$ is the instrumental width, $k_B$ is the Boltzmann constant, $T_{ion}$ is the ion thermal temperature, M is the mass of the ion, and $\xi$ is the non-thermal velocity. The ion temperature in the calculation was replaced with the electron temperature. The main difference from previous study is the height dependence of the electron temperature. Here, the time-dependent ionization effect is considered, and the electron temperature does not change with the height. Figure 3 shows the height dependence of the non-thermal velocity estimated from the line of \ion{Fe}{24} (192.04 \AA) observed by EIS. The diamonds show the same results as those discussed in a previous study. The red solid lines in Figure 3 correspond to the nonthermal velocity at Te = 25 MK, as shown in Figure 2b. The black dashed lines also correspond to those at Te = 20 and 30 K. It can be observed that the non-thermal velocity increases significantly with the height of the sheet structure, even in the case where the time-dependent ionization process is considered. This trend is largely the same in \cite{war}, who did not consider the time-dependent ionization process.

\section{SUMMARY and DISCUSSION}

I have studied plasma heating considering the time-dependent ionization process during a large solar flare on September 10, 2017, as observed by Hinode/EIS. The observed \ion{Fe}{24} / \ion{Fe}{23} ratios are consistent with the results of considering the time-dependent ionization effect at a constant electron temperature Te = 25 MK. It is also discussed that the non-thermal velocity considers the time-dependent ionization process during a large solar flare. The non-thermal velocity increases significantly with the height of the sheet structure, even when considering the time-dependent ionization process.

Let us discuss the validity of our results by comparing them with those of previous studies. As mentioned above, \cite{war} studied the same flare observation data. They found that plasma heating appeared to occur downstream of the reconnection outflow, and that the plasma was heated to temperatures from $\sim$10 to $\sim$25 MK under the assumption of ionization equilibrium. On the other hand, I considered the ionization process and performed an analysis of the reconnection region assuming isothermal temperature. As a result, it was found that the \ion{Fe}{24} / \ion{Fe}{23} observed ratio can be well explained assuming a density of approximately 10$^{10}$ / cc, a velocity of 1000 km s$^{-1}$, and a temperature of 25 MK in the magnetic reconnection region. The density and velocity used in the analysis were higher than those normally observed in the reconnection region. Note that the estimated spatial profiles of \ion{Fe}{24} / \ion{Fe}{23} ratios are the same at a density of 5x10$^9$ /cc and a velocity of 500 km s$^{-1}$. 
\cite{lon} and \cite{che} reported the lower velocity (approximately 200-300 km s$^{-1}$) in the same event. 
In addition, they mentioned that downflows show a rapid deceleration--a behavior that has been well reported in numerous previous studies on SADs velocities. 
At low speeds such as 300 km s$^{-1}$, the plasma in the outflow region may approach ionization equilibrium, and it may be assumed that the temperature of the plasma is not isothermal in the height direction but has a gradient.
In contrast, the velocity at the beginning of the flare ($\sim$16:05) seems to be high-- approximately 500-800 km s$^{-1}$ (see Figure 7 in \cite{lon}).
Therefore, the assumption of a flow velocity of 1000 km s$^{-1}$ is valid in the beginning of the flare.
In fact, \cite{war} studied the temporal evolution of height dependence of the \ion{Fe}{24}/\ion{Fe}{23} ratio, and discovered that the gradient became more gradual over time.
The outflow in the later phase of the flare is slower than the flow in the beginning, and the plasma in the outflow region may approach the ionization equilibrium.
To determine whether there is a temperature gradient in the magnetic reconnection region, it is necessary to perform density, temperature, and velocity diagnostics considering the non-equilibrium ionization. For this purpose, it is necessary to observe other ionization degree lines, for example, from \ion{Fe}{19} to \ion{Fe}{22}. The Solar-C(EUVST) mission currently under consideration has many high-temperature lines and line pairs capable of density diagnosis. The plasma diagnostic studies with regard to the reconnection region are expected to progress dramatically with the Solar-C(EUVST).
From a theoretical standpoint, the temperature downstream of the reconnection region is considered to be almost isothermal until the termination shock, because the energy transport by the thermal conduction is large in the corona. These temperature structures are confirmed by various numerical calculations \citep[e.g.,][]{yok2}. In contrast, there exists theoretically controversy as to whether the method of providing thermal conduction during a flare is appropriate \citep[e.g.,][]{ima2015}, and further discussion is required on this subject.

The effect of non-equilibrium ionization on the line width (nonthermal velocity) was also analyzed. The non-thermal velocities were found to be sufficiently large regardless of whether ionization non-equilibrium was considered. The relationship between plasma heating/acceleration and turbulence, as discussed by  \cite{war} and  \cite{pol} does not change even when non-equilibrium ionization is considered. On the other hand, the energy related to the line width is not very large compared to the energy required to heat the plasma (nonthermal velocity $<<$ thermal velocity), and further discussion on how much it can contribute to plasma heating is necessary.

Finally, I discuss the relationship between the magnetic reconnection model and non-equilibrium ionization. Currently, there are two types of theoretical models of magnetic reconnection: Petschek-type reconnection, which is associated with large-scale slow-mode shocks  \citep[e.g.,][]{yok2}, and plasmaoid-unstable type reconnection in which many plasmoids are formed  \citep[e.g.,][]{bha,shi}. The two models can be distinguished observationally by considering the ionization process in the solar corona. Petschek-type reconnection is usually thought to be accompanied by laminar flow, and ionization is also thought to be a layered structure, as seen in \cite{ima2011c}. In contrast, in the case of plasmoid-unstable reconnection, the heated plasmoids sometimes coalesce and/or wander through the reconnection region to the right and to the left. Therefore, ionization in the reconnection region may not be a layered structure, but a structure in which multiple ionized states coexist (Figure 4). For example, in the reconnection event I analyzed, the reconnection region can be interpreted as the ionization progresses as it goes to the outside of the reconnection region (a layered structure). Hence, it can be considered that Petschek-type reconnection might have occurred in this event. On the other hand, in the limb flare event observed at \cite{tak}, the reconnection region was observed at various wavelengths (emission lines) at the same time and location, and the plasma is considered to be a mixture of multiple ionized states. In addition, many plasmoids were also observed iduring this event. Therefore, it can be considered that a plasmoid-unstable reconnection might occur in that event. The number of high-temperature lines observed by Hinode/EIS is limited, so it is difficult to make a sufficient diagnosis of the reconnection region. 
Recently, the next-generation solar observation satellite Solar-C (EUVST) has been discussed intensively. 
An ultraviolet imaging spectrometer with dramatically improved spatial and temporal resolution is planned for this satellite. 
In the Solar-C era, thermal non-equilibrium plasma will be extensively discussed. 
I expect that Solar-C (EUVST) will reveal the reconnection region in detail.

\acknowledgments The authors thank K. Kusano, T. Kawate, H. Iijima, T. Shibayama for fruitful discussions. This work was partially supported by the Grant-in-Aid for 17K14401 and 15H05816.

\begin{figure}
\epsscale{0.8}
\plotone{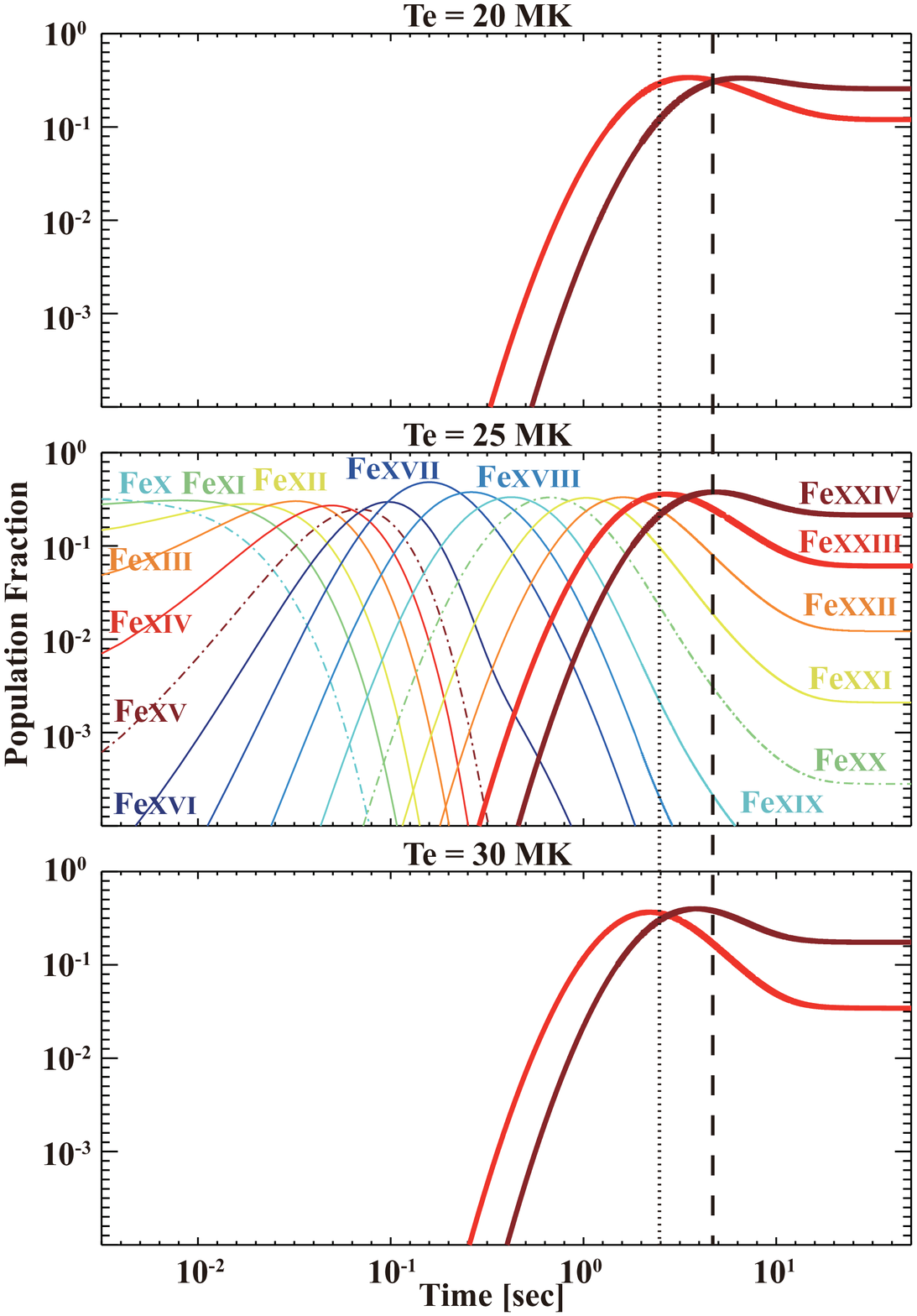}
\caption{
Example of time-dependent ionization with Te = 20, 25, 30 MK. The horizontal axis show the time in logarithmic scale, and the vertical axis show the ionization fraction of iron in logarithmic scale. The dotted/dashed line show the peak time of \ion{Fe}{23}/\ion{Fe}{24}, respectively. The calculation was carried out in the plasma comoving frame.}
\end{figure}

\begin{figure}
\epsscale{0.7}
\plotone{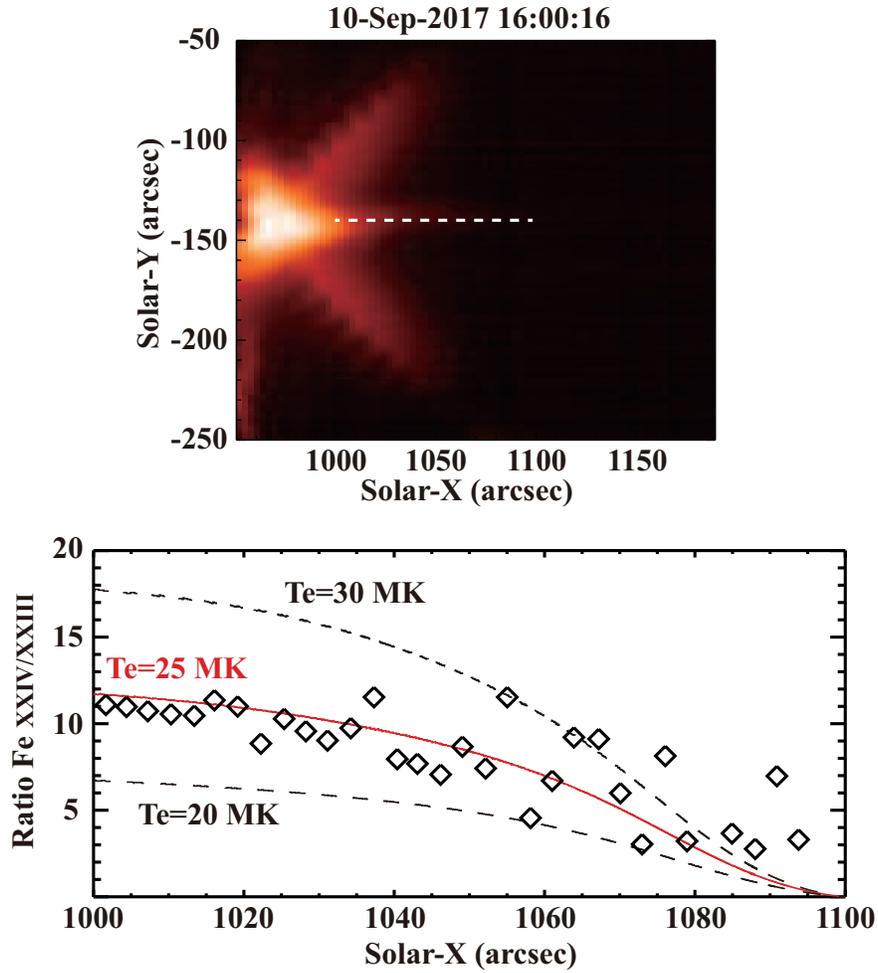}
\caption{
EIS \ion{Fe}{23} (263.76 \AA) image of flare (a), intensity ratio between \ion{Fe}{24} (255.10 \AA) and \ion{Fe}{23} (263.76 \AA) along the downstream of the reconnection outflow (b). Diamonds represent the observation. The red solid line and the black dashed lines represent the theoretical estimation of intensity ratio with constant temperature.}
\end{figure}

\begin{figure}
\epsscale{0.7}
\plotone{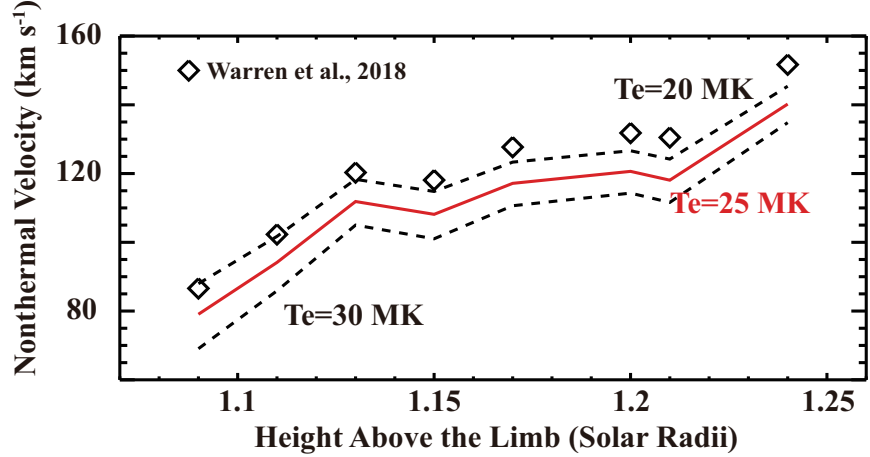}
\caption{
Height dependence of the non-thermal velocity estimated from the line of \ion{Fe}{24}  (192.04 \AA) observed by Hinode/EIS.
}
\end{figure}

\begin{figure}
\epsscale{0.8}
\plotone{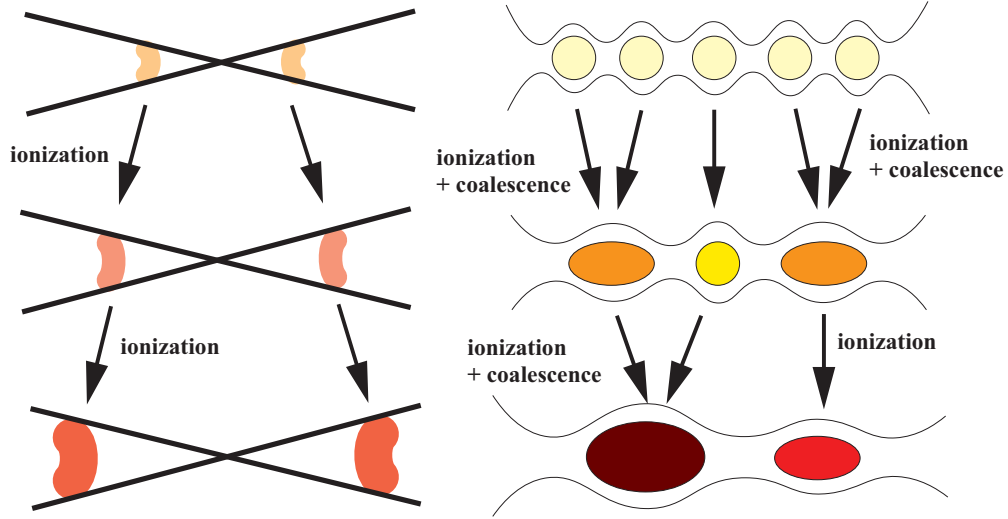}
\caption{ 
Schematic illustration of the ionization in the Petschek-type and plasmoid-unstable reconnection region.
The color depths represent the degree of ionization.
}
\end{figure}

\end{document}